\documentclass[aoas,nameyear,dvips]{arximspdf}
\usepackage{multirow,dcolumn}
\usepackage{graphicx}

\doi{10.1214/10-AOAS349}
\volume{4}
\issue{4}
\pubyear{2010}
\firstpage{1722}
\lastpage{1748}

\makeatletter
\newcolumntype{d}[1]{D{.}{.}{#1}}
\newcommand{\ud}{\,\mathrm{d}}
\newcommand{\TMRCAi}{\mathbf{T}_{i \cdot}}
\newcommand{\TMRCAj}{\mathbf{T}_{\cdot j}}
\newcommand{\TMRCAij}{T_{ij}}
\newcommand{\iTMRCA}{T}
\newcommand{\TMRCAs}{\mathbf{T}}
\newcommand{\IND}{\mathrm{IND}}

\newcommand{\gen}{g_i}
\newcommand{\igen}{g}
\newcommand{\phylo}{\mathbf{Q}_i}
\newcommand{\Ne}{\bolds{\Phi}}
\newcommand{\iNe}{\phi}
\newcommand{\lineages}{k}
\newcommand{\bandwidth}{a}
\newcommand{\numBandwidth}{A}
\newcommand{\seq}{\mathbf{D}_i}
\newcommand{\seqs}{\mathbf{D}}
\newcommand{\MI}{\bolds{\Omega}_0}
\newcommand{\MH}{\bolds{\Omega}}
\newcommand{\nSeg}{I}
\newcommand{\nYears}{J}
\newcommand{\rnYears}{J-1}
\newcommand{\gdr}{\operatorname{diag}(\gamma_1, \ldots, \gamma_{\rnYears})}
\newcommand{\taxa}{N}
\newcommand{\event}{e}
\newcommand{\numEvent}{E}
\newcommand{\isoIntercoalescent}{u}
\newcommand{\group}{S}
\newcommand{\lastEvent}{L}
\newcommand{\firstEvent}{F}
\newcommand{\samplingTimes}{O}
\newcommand{\indexNe}{b}
\newcommand{\numNe}{B}
\newcommand{\indj}{\gamma_j}
\newcommand{\indDiag}{\bolds{\Gamma}}
\newcommand{\iInd}{\gamma}
\newcommand{\timeDesign}{\mathbf{Z}}
\newcommand{\timeDesignj}{\mathbf{Z}_j}
\newcommand{\timeDesignij}{\mathbf{z}_{ij}}
\newcommand{\RE}{\bolds{\beta}}
\newcommand{\REj}{\beta_j}
\newcommand{\iRE}{\beta}
\newcommand{\gvs}{\bolds{\phi}}
\newcommand{\FE}{\bolds{\theta}}
\newcommand{\FEi}{\theta_i}
\newcommand{\iFE}{\theta}
\newcommand{\cov}{\bolds{\Sigma}}
\newcommand{\permute}{\bolds{\pi}_v}
\newcommand{\subgroup}{\tilde{S}}
\newcommand{\subevent}{\tilde{G}}
\newcommand{\subeventc}{\tilde{G}^{c}}
\makeatother

\begin{document}
\begin{frontmatter}

\title{Reuse, recycle, reweigh: Combating influenza through efficient
sequential Bayesian computation for massive data\thanksref{T1}}
\thankstext{T1}{Supported by NIH Grants GM008185, GM086887, and GM053275.}
\runtitle{Reuse, Recycle, Reweigh}

\begin{aug}
\author[A]{\fnms{Jennifer A.} \snm{Tom}\corref{}\ead[label=e1]{jentom@ucla.edu}},
\author[B]{\fnms{Janet S.} \snm{Sinsheimer}\ead[label=e2]{janet@mednet.ucla.edu}}
\and
\author[B]{\fnms{Marc A.} \snm{Suchard}\ead[label=e3]{msuchard@ucla.edu}}

\runauthor{J. A. Tom, J. S. Sinsheimer and M. A. Suchard}
\affiliation{University of California}
\address[A]{J. A. Tom \\
Department of Biostatistics\\
UCLA School of Public Health \\
Los Angeles, California 90095 \\
USA\\
\printead{e1}} 
\address[B]{J. S. Sinsheimer \\
M. A. Suchard \\
Departments of Biomathematics\\
\quad and Human Genetics \\
David Geffen School of Medicine at UCLA\\
Department of Biostatistics \\
UCLA School of Public Health \\
Los Angeles, California 90095\\
USA\\
\printead{e2}\\
\hphantom{E-mail: }\printead*{e3}}
\end{aug}

\received{\smonth{12} \syear{2009}}
\revised{\smonth{3} \syear{2010}}

%
\begin{abstract}
Massive datasets in the gigabyte and terabyte range combined with the
availability of increasingly sophisticated statistical tools yield
analyses at the boundary of what is computationally feasible.
Compromising in the face of this computational burden by partitioning
the dataset into more tractable sizes results in stratified analyses,
removed from the context that justified the initial data collection. In
a Bayesian framework, these stratified analyses generate intermediate
realizations, often compared using point estimates that fail to account
for the variability within and correlation between the distributions
these realizations approximate. However, although the initial
concession to stratify generally precludes the more sensible analysis
using a single joint hierarchical model, we can circumvent this outcome
and capitalize on the intermediate realizations by extending the
dynamic iterative reweighting MCMC algorithm. In doing so, we reuse
the available realizations by reweighting them with importance weights,
recycling them into a now tractable joint hierarchical model. We apply
this technique to intermediate realizations generated from stratified
analyses of 687 influenza A genomes spanning 13 years allowing us to
revisit hypotheses regarding the evolutionary history of influenza
within a hierarchical statistical framework.
\end{abstract}

%

\end{frontmatter}

\section{Introduction}\label{sec:intro}

\subsection{Studying the evolution of influenza A}\label{sec:intro_flu}

Influenza A continues to evade eradication resulting in ongoing
economic and human cost. Yearly epidemics are responsible for 36,000
deaths on average in the United States. Three times in the past
century global pandemics, including the infamous Spanish influenza of
1918, resulted in catastrophic mortality rates [\citet{Salomon09}].
Influenza epidemiologists believe a future influenza pandemic is an
imminent threat [\citet{Webster03}]. Nearly 400 documented transfers
[\citet{Salomon09}] of the highly virulent and potentially pandemic
[\citet{Fauci05}] H5N1 strain of avian flu to humans in addition to the
recent development of H1N1 swine flu [\citet{Butler09}] bolster the
threat. The increasingly relevant necessity of preventing future
influenza pandemics requires a clear understanding of the evolutionary
mechanisms of influenza as it is the key to vaccine development
[\citet{Ghedin05}].

Influenza A is a negative single-stranded RNA virus composed of 8
segments that total approximately 13 kb in length and encode 11
proteins. The three largest segments encode polymerases PB1, PB1-F2,
PB2 and PA all of which are involved in RNA transcription and
replication. The next three segments code for the two surface
glycoproteins haemagglutinin (HA) and neuraminidase (NA) as well as the
nucleoprotein (NP). The two smallest segments encode the matrix
proteins M1 and M2 and the nonstructural proteins NS1 and NS2
[\citet{Yewdell02}, \citet{Nelson07}]. Influenza A research typically focuses
on the epitope-rich HA and NA segments because they exhibit strong
evolutionary selective pressure due to their direct interaction with
the host immune system and are the primary determinants of the
antigenic variation of influenza [\citet{Ghedin05}]. The 16 HA and 9
NA glycoproteins found in the avian reservoir, referred to as H1 to H16
and N1 to N9, respectively, characterize and name the subtypes of
influenza A [\citet{Nelson07}].

The evolutionary history of influenza A involves the interaction of a
number of mechanisms including mutation and reassortment.
Approximately one random sequence mutation every replication cycle
combined with the selective pressure on the surface glycoproteins
results in an accumulation of point mutations on the HA and NA segments
termed antigenic drift. The influenza genome also evolves through
reassortment in which two subtypes coinfect a single host cell and
exchange segments. This exchange of genetic material can lead to an
antigenic shift or the creation of a new, potentially lethal, subtype.
Reassortment between a virus in the avian reservoir and human influenza
A resulted in the subtypes responsible for the Asian and Hong Kong
influenza pandemics [\citet{DeClercq06}] and the current swine flu
pandemic derives from a triple reassortment event [\citet{Smith09}].

As critically important events in influenza evolution occur at the
genome-level, complete genome analysis yields scientific insight that
single segments cannot afford. For example, \citet{Holmes05} clarify a
perplexing question in the evolutionary dynamics of HA by considering
the varying histories of each segment. Analyzing 156 complete H3N2
viruses over a five-year time span from 1999 to 2004,
\citeauthor{Holmes05} (\citeyear{Holmes05}) discover that while the Fujian-variant HA segment
has been co-circulating since at least 2000, the variant only rises to
dominance in 2002 after other segments within the influenza genome
reassort and provide a synergistic background. This important
reassortment event is only understood by studying the influenza genome
in its entirety.

A more recent study by \citet{Rambaut08} emphasizes the importance of
incorporating model parameter uncertainty
in drawing conclusions about influenza evolution through a Bayesian
analysis of a truly massive dataset. \citet{Rambaut08} compile
687 H3N2 influenza A full genomes sampled from New York over a
twelve-year period. \citet{Rambaut08} address a host of
biologically and clinically relevant questions including: (1) Are
reassortment events coincident with shifts in HA antigenicity? (2) Do
certain segments maintain greater genetic diversity? (3) Are the
genetic histories of certain segments correlated? However, due to a
dearth of Bayesian massive data techniques, computational constraints
force \citet{Rambaut08} to partition the data by stratifying on
segment, using the data inefficiently, and drawing ad hoc conclusions
about potential correlation. This current study rectifies the
stratified analyses by fully capitalizing on the hierarchical nature of
the influenza data and making formal inference after modeling the
complete data in a single Bayesian analysis.

\subsection{Statistical context}

Despite optimized algorithms for missing data integration
[\citet{Suchard09}], phylogenetic analysis of DNA sequences is lengthy
and computationally intensive. Massive data measure in the gigabyte to
terabyte range [\citet{Cressie97}] and are increasingly common
[\citet{Lambert03}, \citet{Allison09}]. This pervasiveness is particularly
poignant in Bayesian models with missing data and especially in
Bayesian models for stochastic processes where the dimensionality of
the missing data can far outweigh the observed data. Such is the case
in the evolutionary reconstruction of \citet{Rambaut08}.

One strategy pertinent to massive data inference is stratification
[\citet{Cressie97}, \citet{Kettenring09}], often undesirable because
it comes
shackled with the host of difficulties arising from subgroup analysis
[\citet{Glymour97}, \citet{Lagakos06}]. This identifies the direction that
\citet{Rambaut08} originally follow as they treat each of the
eight segments independently. Shared computer memory and communication
latency between computers limit hopes for considering a proper
hierarchical model across segments simultaneously through which to
share information and learn about segment-to-segment correlation. Even
on state-of-the-art equipment, simulating sufficient realizations from
posterior distributions conditional only on the data from a single
segment, or what we refer to as ``stratified distributions,'' still
compels one to devote weeks of computing time per segment. This huge
computational investment raises a critical point regarding a massive
dataset with hierarchical structure. Often researchers perform
preliminary analysis stratified by the exchangeable identifiers in the
data simply because the statistical tools and computational resources
exist for the stratified analysis. In attempting to fit the full
hierarchical analysis, the ability to reuse the results from these
suboptimal analyses represents a major savings in terms of time and
resources and may even be the only feasible option.

To this end, we examine the dynamic iteratively reweighting MCMC
algorithm (DyIRMA) [\citet{Liang07}, \citet{Liang09}]. DyIRMA is
based on the
meta-analysis technique of using summary statistics from independent
studies to infer a single hierarchical model [\citet{Carlin92},
\citet{Warn02}]. Instead of summary statistics, however, DyIRMA combines
realizations from independent distributions using importance sampling
and Markov chain Monte Carlo (MCMC) in an iterative process.
Importantly, we can adopt DyIRMA to reuse realizations from preliminary
stratified analyses. This desire to not waste intermediate
realizations from the stratified analyses is particularly relevant in
our influenza example because the realizations themselves require
massive computing resources to generate. We further extend the insight
of \citet{Liang07} who combine intermediate realizations from two
uncorrelated distributions. Our extension is necessary to accommodate
correlated sequence data sampled over a span of thirteen years and
allows us to entertain a much richer collection of hierarchical models
motivated by the science at hand.

Our scientific aim in this study is to create a joint hierarchical
model that addresses the questions raised by \citet{Rambaut08}
regarding the evolutionary history of influenza A. To this end, the
hierarchical model must account for an unknown correlation structure
between segments and allow for a flexible time-course in the model
response, for which we exploit Gibbs variable selection (GVS)
[\citet{Kuo98}, \citet{Dellaportas02}] to estimate a nonparametric response.
The influenza A example illustrates that DyIRMA is a particularly
flexible and valuable approach that reuses realizations via reweighting
from computationally expensive distributions in a hierarchical
framework. This widely applicable technique can be used to jointly
model other independently generated, but in truth correlated, massive
datasets.

As a preview, the paper continues as follows:
Section \ref{sec:stratified} describes the generation of realizations
from the stratified analyses, Section \ref{sec:sampling} introduces the
basic framework used to estimate genealogies.
Section \ref{sec:recycle} relates the machinery necessary to combine
these realizations to estimate the joint hierarchical model along with
computational concerns. Section \ref{sec:model} reviews the
hierarchical model proposed, prior distributions, MCMC sampling
concerns, and various modeling extensions.
Sections \ref{sec:theresults} and \ref{sec:discuss} present results and
conclude with a discussion.

\section{Genomic-scale phylogenetic models}\label{sec:genomic}

\subsection{Intermediate phylogenetic realizations} \label{sec:stratified}

Rambaut et~al. (\citeyear{Rambaut08}) compile aligned sequence data for coding regions of
each of the eight
segments of the influenza A genome from the Influenza Genome Sequencing
Project NCBI database [\citet{Ghedin05}]. These alignments contain all
687 H3N2 influenza A genomes available over the 12 influenza seasons
between 1993 and 2005. Season 2002 yields no sequences as it was
predominantly an H1N1 season. From these data, \citet{Rambaut08}
are most interested in estimating and formally comparing the times to
most recent common ancestor (TMRCA) of all the sequences sampled within
each season for each segment. TMRCA can be thought of as a measure of
genetic diversity because evolutionarily distant present-day sequences
converge to a genealogy with longer branch lengths and consequently a
larger TMRCA. To keep computation manageable, \citet{Rambaut08}
are forced to partition these data into independent blocks by segment.
As the virus evolves through time, samples from different seasons are
highly interrelated through their shared history. Standard
phylogenetic software packages account for this correlation. On the
other hand, joint models across segments are less developed
[\citet{Suchard03}]. Consequently, initial analyses consider the
segments independently, clouding conclusions about segment--segment
interactions important to influenza A evolution.

Rambaut's et~al. (\citeyear{Rambaut08}) analyses provide samples from the intermediate
distributions of TMRCAs given sequences from each individual segment.
We are interested in the interaction of the evolutionary dynamics of
influenza A segments over time and we use as our starting point
realizations from these analyses stratified on segment. We let
$T_{ij}$ be the TMRCA for segment $i$ and season $j$ and
$\seq$ the sequence data for segment $i$. From each of the
stratified analyses, we tabulate samples $\{T_{ij}^{(m)} \vert\seq\}$
or $\{T_{ij}^{(1)},\ldots, T_{ij}^{(M)} \vert\seq\}$ for all
$(ij)$ where $m = (1,\ldots,M)$ indexes the MCMC sample, $M$
is the total number of MCMC samples, $i = (1, \ldots, \nSeg)$
indexes the segment, and $j = (1, \ldots, \nYears)$ indexes the
season. Next, let $\TMRCAs^{(m)}$ be the matrix constructed from
$\nYears$ columns $\TMRCAj^{(m)}$ or $\nSeg$ rows $\TMRCAi^{(m)}$,
where $\TMRCAj^{(m)}$ is an $\nSeg\times1$ vector with all samples of
TMRCA at iteration $m$ for season $j$ and $\TMRCAi^{(m)}$ is a
$\nYears\times1$ vector for segment $i$ with all samples of TMRCA
at iteration $m$ for all $\nYears$ seasons. These $M$ matrices
$\TMRCAs^{(m)}$ are the intermediate samples from the stratified
distributions provided by \citet{Rambaut08} that we propose to
recycle into a hierarchical model.

To describe the construction of this stratified distribution, we first
introduce some nomenclature. In brief, let $\igen$ be the genealogy
composed of a bifurcating acyclic graph (commonly called a topology)
that describes the relatedness of a set of sequences and a vector of
edge weights for the edges in this topology. Edges reflect the passage
of time between bifurcation events and are also called branch lengths.
Estimates of $c$ different TMRCA, $\TMRCAs$, embed in $\igen$ because
$\TMRCAs= f(\igen)$ where $f(\cdot)$ is a deterministic mapping$\dvtx g
\rightarrow\Re_{\geq0}^c$. Here, $\TMRCAs$ represents a $c \times1$
vector containing the TMRCAs of interest and we use this general vector
as the starting point before building up to $\TMRCAj$.

In order to describe $f(\cdot)$, a brief introduction to the coalescent
process is warranted [\citet{Hudson91}]. In the isochronous case,
there are $\taxa$ sequences sampled at the same time $t_0$ where $t_0 =
0$ represents the present-day. Formation of a genealogy begins by
randomly selecting two lineages at time $t_\event$, $\event= (1,
\ldots, \numEvent)$ where $\event$ indexes the coalescent event and
$\numEvent= (\taxa- 1)$ is the total number of coalescent events.
Proceeding back in time, the inter-coalescent time between the
$\event$th and $(\event-1)$th event is $\isoIntercoalescent_\event=
t_\event- t_{\event-1}$ where $(\isoIntercoalescent_1, \ldots,
\isoIntercoalescent_{\numEvent})$ are independent exponential random
variables. Let $\group$ be the set of all $\taxa$ taxa; then the
$\numEvent$ independent intercoalescent intervals yield an estimate
$T_\group$, the TMRCA of set $\group$. Since each inter-coalescent
interval is distributed as an exponential, the summation is a
convolution of exponential distributions [\citet{Hein}].

We can estimate $\iTMRCA$ for any subset of taxa. Let $\subgroup
\subseteq\group$ be a subset of the taxa, $\firstEvent$ represent the
first event in this subset which for the isochronous case occurs at
$\firstEvent= 1$, and $\lastEvent$ the last so that $0 < \firstEvent
\leq\lastEvent\leq\numEvent$. Then $T_{\subgroup}$ is calculated
similarly by identifying $\igen_{\subgroup}$, the subtree formed within
$\igen$ from the tips at $t_0$ and proceeding back in time to the last
coalescent event occurring at $t_\lastEvent$ for all taxa in
$\subgroup$. We then calculate $T_{\subgroup} = f(\igen_{\subgroup
}) =
\sum_{\event=\firstEvent}^{\lastEvent}\isoIntercoalescent_\event$.
Finally, as a nomenclature device, let $\subevent$ be the set of all
times of coalescent events identified in $\subgroup$, $\subevent=
\{t_\firstEvent, \ldots, t_\lastEvent\}$, and let $\subeventc$ be the
complement. Then the genealogy can be decomposed into two disjoint
sets of coalescent times as $\igen= (\subevent, \subeventc)$, a device
that will prove useful in Section \ref{sec:reweigh}.

One particularly interesting subset $\subgroup_{j}$ arises in the
case of heterochronous data, namely $T_j$ or TMRCA for all sequences
sampled in a given season $j$. Because we have influenza A
sequences sampled over time, our data are commonly called
``heterochronous.'' We know the exact date of sampling for the
influenza A sequences and can extend the coalescent to reflect this
additional information. Elaborating on the description of the
coalescent process above, the heterochronous case has two events of
interest, coalescent and sampling, both of which can occur at multiple
times. If there are $\samplingTimes$ sampling times, there are now a
total of $(\taxa+ \samplingTimes- 2)$ intervals, so in the
heterochronous case, $\numEvent= (\taxa+ \samplingTimes- 2)$. Let
$(t_0 ,\ldots, t_\numEvent)$ represent the times of coalescent or
sampling events where $t_0$ is the most recent chronological sample and
as before, the inter-event intervals are $\isoIntercoalescent_\event=
t_\event- t_{\event-1}$. For $\subgroup_{j}$ identify the
earliest event $\firstEvent$ and the last event $\lastEvent$ and again
we have $T_{\subgroup_j} = f(\igen_{\subgroup_j}) =
\sum_{\event=\firstEvent}^{\lastEvent} \isoIntercoalescent_\event$.

\begin{figure}

\includegraphics{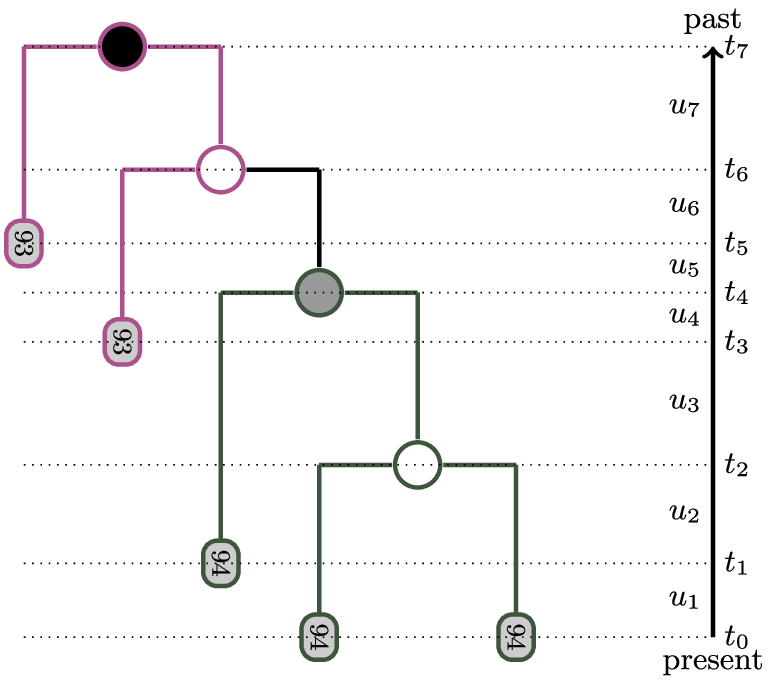}

\caption{\textup{Phylogenetic tree: calculating \iTMRCA, the time to
most recent common ancestor (MRCA).} Five influenza A sequences are
represented by gray rectangles and labeled with the sampling season.
Present time is labeled as $t_0$ and extends back into the past until
$t_7$ or the $T$ for all five sequences. Inferred ancestral nodes
between the samples are represented by circles. The black circle is
the most recent common ancestor for both of the sequences sampled in
the 1993 influenza season. We calculate $T_{1993}$ by isolating the
relevant branches, represented here by the purple subtree, and summing
the inter-coalescent intervals, $T_{1993} = \isoIntercoalescent_4 +
\isoIntercoalescent_5 + \isoIntercoalescent_6 + \isoIntercoalescent_7$.
Similarly, the gray circle is the most recent common ancestor of all
three sequences sampled in the 1994 influenza season and $T_{1994} =
\isoIntercoalescent_1 + \isoIntercoalescent_2 + \isoIntercoalescent
_3 + \isoIntercoalescent_4$.} \label{fig:tree}
\end{figure}

We can now fully describe the intermediate realizations available from
\citet{Rambaut08}. Recall that our research goals require we
extract estimates of $\TMRCAij$ for a specific influenza season
$j$ where $j = (1, \ldots, \nYears)$ indexes the $\nYears$
distinct seasons from a $\gen$ for segment $i = (1, \ldots, \nSeg)$.
We accomplish this by allowing the subset $\subgroup= \group_{ij}$ to
be the set of all taxa for a given influenza season $j$. Then
$\TMRCAij= f(\igen_{i j})$ is the summary statistic of
interest for segment $i$ and season $j$. Finally, let $\TMRCAi
= (\iTMRCA_{i 1}, \ldots, \iTMRCA_{i \nYears})$ contain all
estimates of TMRCA for a given segment and let $\TMRCAj= (\iTMRCA_{1
j}, \ldots, \iTMRCA_{\nSeg j})$ be all estimates of TMRCA for
a given year. Refer to Figure \ref{fig:tree} for a simple example.

\subsection{Estimating TMRCA}\label{sec:sampling}

Although we have outlined how to generate our summary statistics of
interest given a genealogy, further description is necessary regarding
sampling from the distribution of the unknown genealogy $\gen$
conditional on the data $\seq$ for segment $i$. This posterior
distribution can be represented as
\begin{equation}\label{eq:phylo}
P(\gen\vert\seq) \propto\int\int\mathcal{L}(\seq\vert\gen,
\phylo)P(\gen\vert\Ne_{i})P(\Ne_{i})P(\phylo)\ud\Ne_{i}
\ud\phylo,
\end{equation}
where $\mathcal{L}(\seq\vert\gen, \phylo)$ is the likelihood of the
sequence data given the genealogy and other phylogenetic parameters
$\phylo$ that model sequence change over time and $P(\gen\vert
\Ne_{i})$, $P(\Ne_{i})$, and $P(\phylo)$ are the prior
distributions for the genealogy and phylogenetic parameters. The above
decomposition specifies a marginal prior distribution on $\gen$ because
$P(\gen) = \int P(\gen\vert\Ne_{i})P(\Ne_{i})\ud\Ne_{i}$. We
take note of this marginal distribution because we wish to ultimately
replace $\prod_i P(\gen)$ by a joint prior distribution
$P(\igen_1,\ldots,\igen_{\nSeg})$ in our hierarchical model. From
\citet{Rambaut08}, $P(\gen)$ derives from a semiparametric
relaxation of the coalescent process parameterized in terms of
time-varying effective population size vector $\Ne_{i}$ that follows
a piecewise constant multiple-changepoint-process hyperprior
distribution [Drummond et~al. (\citeyear{Drummond02}, \citeyear{Drummond05})]. Effective population
size is meant to reflect amount of genetic diversity rather than census
count [\citet{Wright31}] and can be thought of as the average number of
unique individuals that actually contribute genes to subsequent
generations. Investigators generate samples from $P(\gen\vert\seq)$
using MCMC in the Bayesian software BEAST [Drummond et~al. (\citeyear{Drummond02}, \citeyear{Drummond05})] for each segment independently.

Due to the prominence the prior distribution $P(\gen)$ plays in the
iterative reweighting scheme, it is outlined in some detail as follows,
where we drop the subscript $i$ for clarity. Recall that $(t_0,
\ldots, t_{\numEvent})$ are the times of events going into the past and
$(\isoIntercoalescent_1, \ldots, \isoIntercoalescent_{\numEvent})$ are
the inter-coalescent intervals. Let $(\lineages_1, \ldots,
\lineages_{\numEvent})$ be the number of lineages that exist in
$\igen$
during a given inter-coalescent interval. We want to generate a
sequence of effective population sizes of length $\numNe$ where $1
\leq
\numNe\leq\numEvent$ indexed by $\indexNe= (1, \ldots, \numNe)$ with
time similarly partitioned into $(\tilde{t}_1, \ldots,
\tilde{t}_\numNe)$. Essentially, we want to partition $\Ne$ into
$\numNe$ groups $(\iNe_1, \ldots, \iNe_\numNe)$ where $\phi
_\indexNe$
is constant between $\tilde{t}_{\indexNe}$ and $\tilde{t}_{\indexNe-
1}$. In the heterochronous case, the number of lineages can increase
(for a sampling event) or decrease (for a coalescent event) so there
are two events of interest that can change the number of lineages.
These events are distinguished by the indicator function
$1_{\mathrm{coa}}(\event)$ which indicates that $\event$ is a
coalescent event. \citet{Rambaut08} specify the following
heterochronous semiparametric coalescent prior distribution (again,
ignoring dependence on $i$ for clarity):
\begin{equation}\label{eq:Eq2}
P(g \vert\Ne) = \displaystyle\prod_{\event=1}^{\taxa+
\samplingTimes- 2}\biggl\{\frac{\lineages_\event(\lineages_\event-
1)}{2\phi_{h(\event)}}\biggr\}^{1_{\mathrm{coa}}
(\event)}\exp\biggl(-\frac{\lineages_\event(\lineages
_\event-1)\isoIntercoalescent_\event}{2\phi_{h(\event)}}\biggr),
\end{equation}
where the function $h$ maps from the larger number of $\numEvent$
events to the $\numNe$ groups, or in other words $h(e) = b$ if
$t_\event$ is between $\tilde{t}_\indexNe$ and $\tilde{t}_{\indexNe-
1}$. Finally, to complete the \mbox{specification} of the prior distribution,
the first effective population size follows a scale-invariant prior
distribution, $P(\iNe_1) \propto\frac{1}{\iNe_1}$ [\citet{Drummond05}]
and the remaining $\numNe- 1$ effective population sizes are
distributed as exponential with scale parameter equal to the previous
effective population size, $\phi_\indexNe\sim$
$\operatorname{Exp}(\phi_{\indexNe-1})$. We make no claims about the
appropriateness of this prior distribution choice. However, since the
mean and variance of this prior distribution on $\phi_\indexNe$ grow
with $\indexNe$ and lead to some difficulty later, we point out that
\citet{Minin08} provide a stable alternative with the joint skyride
prior distribution on $P(\gen, \Ne_{i})$.

Now we have a foundation for understanding how genealogies are sampled
and summarized using $\TMRCAij$ and what realizations from the
intermediate distributions of the stratified analyses we have
available. We wish to point out the benefits of the hierarchical model
whose estimation we describe in the next section. These benefits
include shrinkage estimators, a framework for statistical inference
that accounts for correlation between strata, and models based on all
available data. We now delve into how these independently generated
estimates are combined into a joint statistical model reusing the
preliminary realizations.

\section{Computational recycling}\label{sec:recycle}

\subsection{Reweighting realizations}\label{sec:reweigh}
In addition to the stratified realizations, the
process of reweighting the stratified analysis samples of $\TMRCAi$ to
generate the joint hierarchical posterior distribution through DyIRMA
requires $P(\gen)$ and the ability to evaluate the marginal prior
distribution $P(\TMRCAi)$. We first present the DyIRMA machinery and
then comment on how we extend it to accommodate this dataset. We save
computational concerns regarding the calculation of $P(\TMRCAi)$ for
Section~\ref{sec:compute}.

Recall that we employ the following decomposition of genealogy $\gen=
(\subevent, \subeventc)$ which allows us to relate the summary
statistic $\TMRCAi$ to $\gen$ as
\begin{equation}
P(\TMRCAs_{i,\subevent} \vert\seq, \MI) = \int P(\gen\vert\seq,
\MI)\ud\TMRCAs_{i,\subeventc}.
\end{equation}
The symbol $\MI$ identifies the stratified analyses and for notational
consistency can be viewed as the forthcoming hierarchical model
parameters, $\MH$ fixed at an arbitrary value. The subscript
$\subevent$ is dropped in the following equations to simplify notation.

We have intermediate realizations of the multivariate vector of
$\TMRCAi= (\iTMRCA_{i1}, \ldots, \iTMRCA_{i\nYears})$ under the
individual models generated during the initial stratified analysis. We
combine these realizations into a single joint posterior distribution
conditional on all of the sequence data, $P(\TMRCAs\vert\seqs)$ where
$\seqs= (\seqs_1, \ldots, \seqs_\nSeg)$ and $\TMRCAs=
(\TMRCAs_{1\cdot}, \ldots, \TMRCAs_{\nSeg\cdot})$, by preferentially
weighting samples that have a high likelihood under the hierarchical
model relative to the probability of the prior distribution in the
individual models. We make the following assumptions of conditional
independence, namely that given $\TMRCAs$ the sequence data are
independent of the parameters in the hierarchical model, or $P(\seqs
\vert\TMRCAs, \MH) = P(\seqs\vert\TMRCAs)$. We also assume that given
the hierarchical parameters, the TMRCAs $\TMRCAi$, which we treat as
exchangeable units, are independent or in other words we assume
$P(\TMRCAs\vert\MH) = \prod_{i}P(\TMRCAi\vert\MH)$. Given these
assumptions, we can then use the following relationship:
\begin{equation}\label{eq:IRA}
P(\TMRCAs\vert\seqs) \propto\int\displaystyle\prod_{i =1}^{\nSeg}\biggl(P(\TMRCAi\vert\seq, \MI)
\frac{P(\TMRCAi\vert\MH)}{P(\TMRCAi\vert\MI)}\biggr) P(\MH) \ud\MH.
\end{equation}
Therefore we reuse all available $M$ realizations from $P(\TMRCAi\vert
\seq, \MI)$ by identifying the following importance weights which are
calculated for every sample of the stratified distributions:
\begin{equation}
w(\TMRCAi, \MH) = \frac{P(\TMRCAi\vert\MH)}{P(\TMRCAi\vert\MI)}.
\end{equation}
The numerator of the weights, $P(\TMRCAi\vert\MH)$, is the conditional
density under the hierarchical model. These hierarchical parameters
are updated during each round of Gibbs sampling requiring recalculation
of the weights for each iteration. The proposal density, $P(\TMRCAi
\vert\MI)$, is conditional on the parameters in the stratified
analyses. As the variance of the estimate in (\ref{eq:IRA}) relies on
the proposal density, some thought should be given toward selecting an
appropriate prior distribution during individual analysis. Generally
it is desirable to have a proposal density with heavier tails than the
numerator of the weights [\citet{Robert}].

As shown by \citet{Liang07} and relying on the concept of importance
sampling [\citet{Rubin88}], we can solve for the conditional
distribution of the $\TMRCAi$'s under the hierarchical model, which is
specified as the following:
\begin{equation}
P(\TMRCAi\vert\seqs, \MH)_{\mathrm{DyIRMA}} =
\frac{1}{W_i}\sum_{m=1}^Mw\bigl(\TMRCAi^{(m)}, \MH\bigr)
\delta_{\TMRCAi^{(m)}}(\TMRCAi), \label{eq:DyIRMAcond}
\end{equation}
where $W_i = \sum_{m=1}^M w(\TMRCAi^{(m)}, \MH)$ and
$\delta_{\TMRCAi^{(m)}}(\TMRCAi)$ is a degenerate distribution at
$\TMRCAi^{(m)}$. The weighted stratified realizations are sampled
during each round of Gibbs updates to generate samples from the
hierarchical posterior distributions of $\TMRCAi$ conditional on the
current values of the parameters in the hierarchical model, $\MH=
\MH^{(m)}$. These samples in turn are then used to update the
parameters in the hierarchical model from $\MH^{(m)}$ to $\MH^{(m+1)}$
which completes another iteration of sampling and contributes the last
remaining piece of machinery necessary to jointly model our
intermediate realizations from stratified analyses.

\subsection{Practical computational concerns}\label{sec:compute}

We now describe in detail the strategy we employed
to calculate $P(\TMRCAi)$ which is necessary for the calculation of
weights. We include this description because for many scientifically
interesting choices of the mapping $f(\cdot)$, such as jointly modeling
TMRCAs, $P(\TMRCAi)$ remains intractable in analytic form.
Fortunately, standard machinery already exists to draw simulants from
this distribution, namely the MCMC sampler exploited to generate the
stratified samples. A single additional run of this sampler without
data provides all ingredients necessary to tackle the seemingly
computationally intractable joint inference. Each partitioned dataset
assumes identical prior distributions so a single simulation of
$P(\TMRCAi)$ suffices for the reweighting of all the stratified
distributions.

Two successful approaches to estimating marginal distributions
from\break MCMC samples are multivariate kernel density [\citet{Scott92},
\citet{Cacoullos64}] and importance weighted marginal density estimation
(IWMDE) [\citet{Chen94}]. IWMDE proceeds by identifying a weighting
function and sampling from a weighted ratio of the likelihood at a
given value of the marginal of interest (here $\iTMRCA_{\subevent} =
t_{\subevent}$) and realizations of the full joint distribution. The
weighting function ideally has a similar shape to the unknown
conditional marginal density, a Catch-22 circumvented with multivariate
kernel density estimation. As a consequence of the strictly
nonnegative support of $\TMRCAi$ we control for potential boundary
effects during kernel density estimation by using a gamma kernel
[\citet{ChenSX00}] that removes the boundary effect and has the best
mean integrated squared error among all nonnegative densities. For
computational ease, we select the bandwidth using a multivariate
adapted Scott's rule-of-thumb in which the bandwidth for the
$\bandwidth$th variable, $\bandwidth= (1, \ldots, \numBandwidth)$, is
$M^{{1}/{(\numBandwidth+4)}}\hat{\sigma}_\bandwidth$
[\citet{Hardle04}] where $\hat{\sigma}_\bandwidth$ is the univariate
sample standard deviation and bandwidths are allowed to vary for each
univariate kernel resulting in a multiplicative kernel
[\citet{Hardle90}].

The importance of selecting an appropriate prior distribution in a
Bayesian framework is a topic of considerable depth [\citet{Gelman04},
\citet{Efron86}]. Even the less contentious tactic of selecting a
noninformative prior distribution by placing a uniform distribution
over the parameter space can prove to be subjective [\citet{Kass96},
\citet{Zwickl04}]. In the case where the likelihood function
overwhelms the
prior distribution, a potentially informative uniform prior
distribution is rendered noninformative and specification of the prior
distribution is deceptively unimportant. When \citet{Rambaut08}
specify a multiple changepoint prior distribution, the effective
population sizes, $\Ne_i$, were further constrained to lie between 0
and 120,000. This truncated prior distribution is well outside of the
desired range of the values of $\TMRCAi$, a fact that is
inconsequential when sampling from the distribution conditional on the
data. However, this creates difficulties for evaluating the prior
distribution at the realized values of the distribution conditional on
the stratified data as they mostly lie well outside the region of the
kernel density estimate. Luckily, when the maximum population size is
constrained to be lower than 120,000 and coverage extends to the region
of interest the KDE of the prior distribution reveals a relatively flat
density surface.

\begin{figure}

\includegraphics{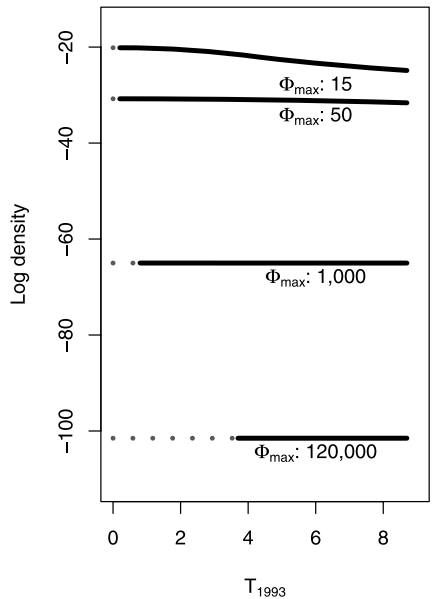}

\caption{Prior density predicted by kernel density estimation (KDE)
under different constraints on the prior distribution for a single
representative season (1993). A prior density predicted by KDE with a
gamma kernel conditional on the mean of the other eleven dimensions
$T_{1994}, \ldots, T_{2005}$. We want the value of the prior density
used during stratified analyses for the intermediate realizations
sampled in the range illustrated in gray dot and black line. We have a
KDE constructed with coverage from realizations of the prior
distribution we generated in the range in black line. The gray dot
region, or the region of interest the KDE is forced to extrapolate,
expands as the maximum allowable $\Ne$, $\Ne_{\max}$,
increases. The density with $\Ne_{\max} = $120,000 approaches
the flat line illustrated with the perhaps more reasonable
$\Ne_{\max} = 15$.} \label{fig:kde}
\end{figure}

As the maximum is gradually increased to 120,000, this surface
decreases in the value of the density but remains relatively flat.
This is illustrated in Figure \ref{fig:kde}, which shows a
representative ($\iTMRCA_{1993}$) prior density predicted by KDE
conditional on the mean of the other parameters for different maximum
allowable $\Ne_i$ sizes along with realizations from the prior
distribution. The range of the values for the realizations from the
prior distribution we wish to evaluate are indicated in the lower range
of the density and the realizations used to predict the KDE are in the
upper range. Taken together, this indicates the region the KDE is
forced to extrapolate. As the addition of a constant value on the log
scale does not affect the weights, a constant density equivalent to
some arbitrary $\varepsilon$ can be selected. For computational ease it
is desirable for $\varepsilon$ to be similar in range to the density under
the joint hierarchical model.
\eject

\section{Hierarchical model: Antigenic shifts and diversity through time}\label{sec:model}
Our methods enable us to conceive of a model that tests a greater range
of hypotheses than those based on a single stratum. We revisit each of
the statements of \citet{Rambaut08} aimed at understanding the
evolutionary history of influenza A with the advantages afforded by a
hierarchical framework. We construct our model out of three basic
elements. The first modeling element identifies seasons with a
significant change in TMRCA from the previous year using GVS on the
timepoints. A~significant increase in TMRCA between timepoints
suggests a reassortment event, whereas a significant decrease suggests
a selective sweep. The second element introduces fixed segment effects
that test whether certain segments have a higher TMRCA and therefore
greater genetic diversity than others across time. Finally, we address
correlation between the segments by exploring constrained variance
matrices. To recycle random samples generated under existing
stratified analyses, we implement an additional DyIRMA step during each
round of Gibbs sampling of the joint model parameters. We first build
up to the biologically motivated mean structure of the model containing
GVS and fixed segment effects with an independent variance matrix and
then introduce extensions for modeling the variance structure.

\subsection{Flexible modeling of time course through Gibbs variable selection}
At the most basic biological level, we must identify significant
changes in TMRCA between influenza seasons and test if these correspond
with shifts in HA antigenicity. We accomplish this with GVS as
parameterized by \citet{Kuo98} where the outcomes of the regression are
the TMRCAs and season effects represent potential predictors. Note
that this analysis would not be possible using a single segment as
there would be insufficient degrees of freedom. The goal of Bayesian
stochastic search variable selection is to identify the underlying
generative model $M_0$ from the set of all possible models $M$. If $J$
is the total number of possible predictors in the regression model, the
model space has dimension $2^{J}$, an arduous dimension from which to
draw inference. \citet{Kuo98} bypass this task by introducing
indicator variables, $\indj\sim\operatorname{Bernoulli}(p_{j})$, that
identify the potential predictors of the outcome variable. To clarify,
if the $j$th predictor has a marked effect on the outcome, the
posterior probability that the corresponding indicator variable is one,
$P(\gamma_j = 1 \vert\seqs)$, is high. On the other hand, if the $j$th
predictor is not critical, the posterior probability that the
corresponding indicator variable is zero, $P(\gamma_j = 0 \vert\seqs)$,
is high. This implies that estimates of $\indj$ clarify which
timepoints correspond to significant changes in TMRCA.

In order to proceed, we must introduce some additional nomenclature.
Let $\timeDesign$ be the $(\nSeg\nYears) \times(\rnYears)$ additive
design matrix for all seasons where we remove the intercept in order to
avoid overparameterization. In other words, $\timeDesignij^{\prime}$
is a row vector where the first $(j-1)$ entries are 1 and the
remaining $(\nYears- j)$ are 0's. Then $\timeDesignj$ is made of
$\nSeg$ identical rows of $\timeDesignij^{\prime}$ and stacking the
$\nYears$ matrices of $(\mathbf{Z}_1, \ldots, \mathbf{Z}_\nYears)$
generates $\timeDesign$. Let $\indDiag$ equal the diagonal matrix
$\gdr$ and $\RE= (\iRE_1, \ldots, \iRE_{\rnYears})^{\prime}$ be the
unknown season effect sizes. The flexible GVS-induced mean time-course
for MRCAs for all segments at season $j$ becomes
$\timeDesignj\indDiag\RE$ and is identical for all segments $i$. We
select the following conjugate, independent, and noninformative priors
distributions: $\REj\sim\mathrm{N}(\mu_{\iRE}, \tau^{-1}_\iRE)$ and
$\indj\sim\operatorname{Bernoulli}(p_{j0})$ where $\mu_{\iRE},
\tau_{\iRE}$, and $p_{j0}$ are hyperprior constants. Estimation of
$\indj$ addresses questions about the evolution of influenza data over
time and clarify whether shifts in HA antigenicity correlate with
significant changes in TMRCA.

\subsection{Modeling the segment effect}
Modeling the segment effect highlights the importance of jointly
modeling the influenza genome in concert in order to draw meaningful
inference about similarities and differences in their evolutionary
histories. Segment effects identify consistent differences in TMRCA
over time and can test the hypothesis that NP has higher genetic
diversity than HA. We also garner indirect information regarding the
unresolved physical location of segments within the influenza A genome
because we can resolve the correlation between segments and highly
correlated segment histories are consistent with close proximity.
Multidimensional scaling (MDS) suggests a relationship with decreasing
intensity among the following three groupings: (1) \{HA, M1$/$2\}, (2)
\{NS1$/$2, NP\}, and (3) \{PA, PB1, PB2\} [\citet{Rambaut08}]. The NA
segment is not grouped with any other segments. Let $\FE= (\iFE_1,
\ldots, \iFE_\nSeg)$ be unknown segment effects. Then we model
\begin{equation}
\TMRCAj\sim\mathrm{N}(\FE+ \timeDesignj\gvs, \cov),
\end{equation}
where in its most general form $\cov$ is assumed to be an $\nSeg
\times
\nSeg$ unstructured (UNS) covariance matrix. Conjugate prior
distributions for this portion of the model are $\FEi\sim
\mathrm{N}(\mu_{\iFE}, \tau^{-1}_{\iFE})$ for all $i$ and $\cov^{-1}
\sim\operatorname{Wishart}(\nu,\mathbf{R}^{-1})$ where $\mu_{\iFE},
\tau^{-1}_{\iFE}$, and $\nu$ are hyperprior constants and
$\mathbf{R}^{-1}$ is the inverse of the hyperprior constant $(\nSeg
\times\nSeg)$ scaling matrix. When combined with the prior
distributions above, drawing realizations from the conditional
posterior distributions for $\cov, \FE, \RE$, and $\indDiag$ helps
address questions about segment effects, significant timepoints, and
segment correlation simultaneously.

\subsection{Sampling from the complete model}
\label{sec:complete} We now specify how to draw MCMC samples from the
complete model. Let $\MH= (\FE, \cov, \indDiag, \RE)$ contain the
unknown parameters of the hierarchical model, of which we wish to draw
inference. Recall that we specify the conditional distribution for
$\TMRCAi$ in (\ref{eq:DyIRMAcond}), which shows how we obtain
realizations from $P(\TMRCAs\vert\MH, \seqs) =
\prod_{i=1}^{\nSeg}P(\TMRCAi\vert\MH, \seqs)$ by reweighting samples
from $P(\TMRCAi\vert\seqs, \MI)$. Therefore to estimate the
parameters from this complete model using Gibbs sampling, we use DyIRMA
during each Gibbs cycle over $(\TMRCAs, \FE, \cov, \indDiag, \RE
)$. We
have stated the denominator of the weights previously as the
predictions from the KDE of the stratified prior distributions that do
not depend on $\MH$ and hence are constant during MCMC sampling. The
numerator is updated at each iteration of MCMC and is simply the
density of the vector of TMRCAs given the parameters in the
hierarchical model, a straightforward way to determine multivariate
normal. At this stage we are replacing the standard coalescent prior
distribution with a normal prior distribution, a trade-off that yields
straightforward interpretability as it allows us to directly test
\citet{Rambaut08} hypotheses, in addition to a simple
computational implementation. For details on the other update steps
for $(\FE, \cov, \RE, \indDiag)$ and a schematic of sampling refer to
the supplementary material [\citet{Tom10}].

\subsection{Modeling extensions}
\subsubsection{Constrained covariance matrices}\label{sec:constrained}
To identify segments with similar evolutionary
histories, several constraints to $\cov$ may provide more effective
estimates. We explore an independent (IND) parameterization, such that
$\cov= \sigma^2\textbf{I}_{\nSeg}$, with marginal variance $\sigma^2$
unknown that implies the evolutionary history of segments is not
correlated. This specification allows inference to focus on the
segment effects and has the additional advantage of substantially
decreasing the number of inferred parameters. Also informative is
compound symmetry (CS) which gives a general estimate of correlation
between segments and provides a model nested within UNS to test for
similar levels of correlation between segments. CS implies that the
evolutionary histories of all segments are correlated with the same
strength. Finally, autoregressive first order (AR1) and tridiagonal
(TRI) structures with an estimable ordering of the segments directly
identifies which segments have similar evolutionary histories. The
motivation for nonexchangeable structures relies on the reasoning that
segments with similar evolutionary histories have higher correlation
than those with dissimilar histories.

For the CS model, we modify the Gibbs sampling by replacing the
step for $\cov$ with a Metropolis--Hastings step. Let $\cov =
\Psi(\boldsymbol{\xi}) = \Psi(\sigma^2, \rho)$ where $\rho$ is the
segment correlation and assume prior distributions $\sigma^2 \sim
\operatorname{Inverse\mbox{-}Gamma}(\alpha_{\sigma^2},\break \lambda^{-1}_{\sigma
^2})$ and
$\rho\sim \operatorname{Beta}(\alpha_{\rho}, \lambda^{-1}_{\rho})$ where
$\alpha_{\sigma^2}, \lambda^{-1}_{\sigma^2}, \alpha_{\rho}$, and
$\lambda^{-1}_{\rho}$ are hyperprior constants. Refer to the
supplementary material [\citet{Tom10}] for further details on this
modification to sampling.

\subsubsection{Finding the optimal correlation between segments}
Segments are not exchangeable in the TRI and AR1 parameterizations of
the covariance matrix. In AR1, nearest-neighbor segments in the
covariance matrix have higher correlation than those further apart. In
TRI, the structure is more restrictive with segments more than one
neighbor away from each other having no correlation. As the ordering
of the segments $i$ is not known and of paramount scientific
interest, we estimate the labeling or ordering of the segments within
the covariance matrix by parameterizing a permutation vector $\permute$
where $v = (1, \ldots, \nSeg!)$ indexes the different possible
permutations. Sampling $\permute$ requires an additional
Metropolis--Hastings step. We propose $\permute^{*}$ by randomly
swapping two positions in the current permutation $\permute$. We
accept this proposal with probability $\min(1,
\alpha_{\permute})$ where $\alpha_{\permute}$ is
\begin{equation}
\alpha_{\permute} = \frac{P(\permute^{*})P(\TMRCAs\vert\permute^{*},
\RE, \FE, \indDiag, \cov, \seqs) }{P(\permute)P(\TMRCAs\vert
\permute,
\RE, \FE, \indDiag, \cov, \seqs)}.
\end{equation}
Assuming all permutations occur with equal probability, the terms
$P(\permute^{*})$ and $P(\permute)$ cancel. We proceed with caution
when drawing inference due to label switching [\citet{Celeux00}].
Instead of focusing inference on the permutation itself, we concentrate
on the posterior probability that a given pair of segments are nearest
neighbors, suggesting these segments have similar evolutionary
histories. Our reasoning is that because only nearest neighbors are
correlated, segments with similar evolutionary histories have a high
posterior probability of adjacent positions within the permutation
vector conditional on the correlation $\rho$ being positive. This
model that includes estimation of $\permute$ (TRI-P) clarifies which
evolutionary histories of segments are strongly correlated, which in
turn has implications regarding structural interaction.

\section{Results}\label{sec:theresults}
Rambaut et al. (\citeyear{Rambaut08}) have run stratified analyses for
each of the eight influenza segments. Each analysis required an
exhausting 2--3 weeks on high-end computers to approximate the
stratified distributions $P(\gen\vert\seq)$ via MCMC. Each MCMC
chain runs for $10^8$ iterations and subsampling every $10^5$
iterations yields $10^3$ approximately independent samples from these
stratified distributions. We recycle these precomputed random samples
to fit our hierarchical model that corrects for the stratification
using the Bayesian machinery described in Sections \ref{sec:genomic}
and~\ref{sec:recycle}. In particular, we implemented the DyIRMA Gibbs
sampling scheme in cross-platform Java. We simulate three independent
MCMC chains for each hierarchical model for $10^6$ iterations with 10\%
burn-in and a 10-fold thinning. Each chain takes only approximately
five hours to run on a mid-end desktop computer, representing a
compelling and efficient alternative to fitting a joint hierarchical
model starting from the sequence data. We assess the combined chains
from three independent chains via several convergence criteria
including trace plots, histograms, Geweke's convergence diagnostic
[\citet{Geweke92}], and Rhat [\citet{Gelman04}].

Table \ref{tab:TMRCA} presents mean time-course estimates for the
across-segments independent (IND) and tridiagonal with permutation
(TRI-P) models. An additional subscript of IND or TRI-P clarifies the
model for the parameter estimate. The additive parameterization of the
design matrix results in parameter estimates that reflect an increase
or decrease from the previous season. Table \ref{tab:TMRCA} therefore
reports both the relative posterior conditional mean of $\REj$,
$E(\REj
\vert\seqs, \indj= 1)$, that reflects the relative change in TMRCA
(from the previous season) and, what we term, the absolute posterior
conditional mean or posterior conditional mean of the segment effect
average added to the cumulative posterior mean of $\REj$,
$E(\iFE_{\cdot} + \sum_{j=1994}^J\REj\vert\seqs, \indj= 1)$ where
$\iFE_{.} = \frac{1}{\nSeg}\sum_{i=1}^{\nSeg}\FEi$. Additionally, we
report the posterior mean estimates for $\indj$ that reflect the
posterior probability that $\REj$ is included in the model.

\begin{table}
\caption{\textup{Posterior estimates for parameters summarizing
time-course via Gibbs variable selection.} We report both the
relative, $E(\REj\vert\seqs, \indj= 1)$, and absolute,
$E(\iFE_{\cdot} \vert\seqs) + E(\sum_{j=1994}^J\REj\vert\seqs,
\indj
=1)$ where $\iFE_{.} = \frac{1}{\nSeg}\sum_{i=1}^{\nSeg}\FEi$, values
of TMRCA. Additive change is captured by posterior mean values for
$\REj$ conditional on selection in a given iteration for both the
independent (IND) and tridiagonal with permutations (TRI-P) models.
The superscript $\dag$ indicates influenza seasons during which there
was an HA antigenic shift} \label{tab:TMRCA}
\begin{tabular*}{\textwidth}{@{\extracolsep{\fill}}ld{1.6}d{2.6}d{1.2}@{}}
\hline
&&\multicolumn{1}{c}{\textbf{Relative posterior}}&\multicolumn{1}{c@{}}{\textbf{Absolute posterior}}\\
\textbf{Year} &\multicolumn{1}{c}{$\bolds{P}\bolds{(}\bolds{\gamma}_{\bolds{j}}\bolds{=} \mathbf{1} \bolds{\vert} \seqs\bolds{)}$}& \multicolumn{1}{c}{\textbf{mean TMRCA}} & \multicolumn{1}{c@{}}{\textbf{mean TMRCA}}\\
\hline
\multicolumn{2}{@{}l}{\textit{Model---Independent (IND)}} &  &  \\
1994 & 0.143 & 0.166 & 2.24 \\
1995 & 0.113 & 0.110 & 2.35 \\
1996 & 0.00667 & 0.00327 & 2.35\\
$1997^{\dag}$ & 0.00115 & 0.0104 & 2.34 \\
$1998^{\dag}$ & 0.0748 & -0.0665 & 2.30 \\
1999 & 0.0207 & -0.0128 & 2.28 \\
2000 & 0.0133 & -0.00605 & 2.27\\
2002 & 0.207 & -0.242 & 2.03 \\
2003 & 1.00 & 3.02 & 5.05 \\
$2004^{\dag}$ & 0.00852 & 0.00538 & 5.05 \\
$2005^{\dag}$ & 1.00 & -3.86 & 1.20 \\[5pt]
\multicolumn{2}{@{}l}{\textit{Model---Tridiagonal (TRI-P)}} & & \\
1994 & 0.0500 & 0.0568 & 2.11 \\
1995 & 0.0356 & 0.0326 & 2.14 \\
1996 & 0.00593 & 0.00235 & 2.14 \\
$1997^{\dag}$ & 0.00556 & 0.00122 & 2.14 \\
$1998^{\dag}$ & 0.00100 & -0.00511 & 2.14 \\
1999 & 0.00815 & -0.00346 & 2.13 \\
2000 & 0.00852 & -0.00268 & 2.13 \\
2002 & 0.0437 & -0.0456 & 2.09 \\
2003 & 1.00 & 2.89 & 4.96 \\
$2004^{\dag}$ & 0.0159 & 0.0118 & 4.98 \\
$2005^{\dag}$ & 1.00 & -3.79 & 1.18\\
\hline
\end{tabular*}
\end{table}

Looking first at the IND model, an indicator estimate of 1 at season
2003 suggests that there exists a significant difference between the
average segment time to MRCA, $\iTMRCA$, at influenza season 2002 and
that of 2003 because the posterior probability that the inclusion of a
regression parameter captures this difference approaches 1. This
decisive support of a significant jump in $\iTMRCA$ suggests an
infusion of genetic diversity and is consistent with a
\textit{reassortment event}. Similarly, the posterior probability for
the indicator representing the shift between seasons 2004 and 2005 also
approaches 1. However, in this instance, as the negative sign of the
estimate for $\iRE_{\mathrm{IND,2005}}$ reflects, this
suggests a decrease in genetic diversity which is consistent with a
\textit{selective sweep}.

\begin{figure}

\includegraphics{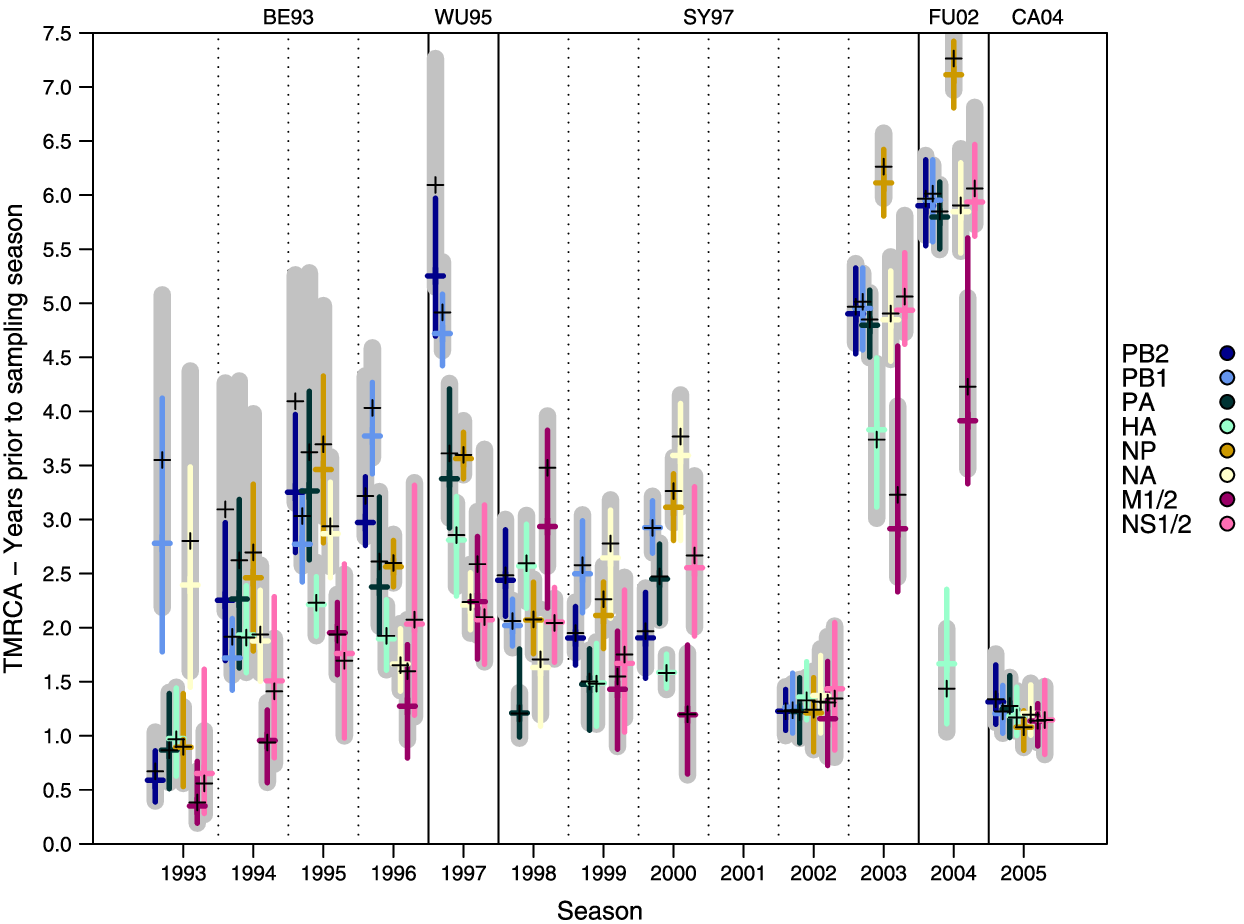}

\caption{\textup{Realizations from stratified and hierarchical
analyses.} Stratified (gray) and hierarchical (colored line) sample
means and 95\% highest density intervals conditioned on data for the
eight segments (PB2, PB1, PA, HA, NP, NA, M1$/$2, and NS1$/$2) of H3N2
influenza A listed from largest to smallest. Samples from the
hierarchical posterior distributions are from the tridiagonal with
permutations (TRI-P) model. Twelve seasons are depicted with season
2001 missing due to an H1N1 dominant season. Segment distributions are
staggered for clarity. The y-axis represents TMRCA and HA antigenicity
(BE93, WU95, SY97, FU02, and CA04) is indicated across the top.
Hierarchical posterior distributions exhibit shrinkage toward the mean
relative to the stratified. }\label{fig:fitted}
\end{figure}

Looking more closely at the IND model, the 2003 shift corresponds to an
increase in $\iTMRCA$ with a posterior mean regression parameter
estimate of $\hat{\iRE}_{\mathrm{IND,2003}} = 3.02$ and
95\% Bayesian credible interval (BCI) $(2.23, 4.28)$. The 2005 shift
suggests a $\hat{\iRE}_{\mathrm{IND,2005}} = -3.86$ $(-4.78, -3.00)$ decrease in $\iTMRCA$ and is concomitant with the
FU02-CA04 HA antigenic shift. No other indicators have posterior
probabilities greater than 0.95. Similarly, the TRI-P model furnishes
strong support for including seasons 2003
($\hat{\iInd}_{\mathrm{TRI\mbox{-}P,2003}} = 1$) and 2005
($\hat{\iInd}_{\mathrm{TRI\mbox{-}P,2005}} = 1$) and shows the
similar pattern of jump in $\iTMRCA$ of $\hat{\iRE}_{\mathrm{TRI\mbox{-}P,2003}} = 2.89$ $(2.00, 3.91)$ followed by decrease of
$\hat{\iRE}_{\mathrm{TRI\mbox{-}P,2005}} = -3.79$ $(-5.03,
-2.55)$. The increase in genetic diversity at season 2003 followed by
a decrease at 2005 clearly identify themselves in the third column of
Table \ref{tab:TMRCA} that shows that the posterior conditional segment
means for the seasons between 1994 and 2002 as around 2 but jumps to
nearly 5 in seasons 2003 and 2004 before undergoing a decline in 2005
and decreasing to around 1.2. Figure \ref{fig:fitted} reiterates this
finding. In general, the TRI-P indicator probabilities are closer to 0
or 1 than in the IND model. For example, the posterior probability of
the season 2002 shift $(\hat{\iInd}_{\mathrm{TRI\mbox{-}P,2002}} =
0.0437)$ is less than that for the IND model
$(\hat{\iInd}_{\mathrm{IND,2002}} = 0.207)$. One
explanation for the differences between the two models is that the
parameter estimate standard errors are generally reduced for the TRI-P
model.

There exists a wide number of covariance structures that are
biologically interesting to explore within our framework.
Specifically, we also consider the unstructured (UNS) and first order
autoregressive with permutation (AR1-P) models. Although we observe a
similar pattern of increase and decrease in TMRCA in 2003 and 2005,
these models that allow a high level of correlation between the
segments lose identifiability of the indicator variables across time,
presumably because the variation present in the observations is used to
model the correlation instead of the mean structure. Combined with the
fact that inference focuses on the mean structure, we continue to
exclusively discuss results for sparse covariance matrices such as
those used in the IND and TRI-P models.

Summarizing the posterior probability of a model that contains all of
the antigenic shifts addresses our question regarding whether HA shifts
are concomitant with significant changes in \textit{T}. Antigenic shifts
occur at 1997 (BE93-WU95), 1998 (WU95-SY97), 2004 (SY97-FU02), and 2005
(FU02-CA04) and these seasons are superscripted in
Table \ref{tab:TMRCA}. We refer to these seasons from now on as shifts
1, 2, 3, and 4, respectively, and use these shift numbers as subscripts
on a given model $M$ to clarify which model variables are included.
For example, $M_{13}$ describes a model where both the first and third
antigenic shifts are included $(\hat{\iInd}_{1997} = 1$ and
$\hat{\iInd}_{2004}=1)$. As we have already noted, the posterior
probability that the FU02-CA04 shift in 2005 is included in both the
IND and TRI-P models approaches 1. However, the posterior probability
of this being the only antigenic shift selected is also large because
$P_{\IND}(M_{4} \vert\seqs) = 0.91$ and $P_{\mathrm{TRI\mbox{-}P}}(M_{4} \vert
\seqs) =
0.97$. The posterior probability of a model with all four antigenic
shifts, $P(M_{1234} \vert\seqs)$, approaches zero for both TRI-P and
IND. This strongly suggests that HA antigenic shifts are not
strictly concomitant with significant changes in TMRCA from the
previous season after correcting for the correlation structure of the
segments. \citet{Rambaut08} are unable to correct for the
correlation structure across segments. Given the limited time series
data available relative to the number of antigenic shifts, more data
would certainly enhance the understanding of the relationship between
reassortment events, selective sweeps, and HA antigenic shifts.

\begin{table}
\tablewidth=310pt
\caption{\textup{Posterior mean and 95$\%$ Bayesian credible intervals
(BCIs) for segment-specific effects.} Posterior mean and probability
intervals for segment effects in both the independent (IND) and
tridiagonal with permutations (TRI-P) models. The superscript $\dag$
indicates our comparison of particular interest. The posterior
probability of NP being greater than HA is 0.995 in the IND model
$(\mathit{BF}_{\IND} = 179)$, assuming equal prior probability. Similarly, for
the TRI-P model, the posterior probability is 0.941 $(\mathit{BF}_{\mathrm{TRI\mbox{-}P}} =
16.1)$. Both provide strong evidence that the evolutionary history of
NP has greater diversity than that of HA} \label{tab:fixed}
\begin{tabular*}{310pt}{@{\extracolsep{\fill}}ld{1.2}d{2.9}@{}}
\hline
&\multicolumn{1}{c}{\textbf{Posterior mean}} & \multicolumn{1}{c@{}}{\textbf{95$\bolds{\%}$ BCI}} \\
\hline
\multicolumn{2}{@{}l}{\textit{Model---Independent}} & \\
PB2 & 2.51 & (1.30,\  3.40) \\
PB1 & 2.65 & (1.42,\ 3.49) \\
PA & 2.11 & (0.888,\ 2.96) \\
$\mathrm{HA}^{\dag}$ &1.37 & (0.161,\ 2.17) \\
$\mathrm{NP}^{\dag}$ & 2.53 & (1.30,\ 3.36) \\
NA & 2.21 & (1.01,\ 3.02) \\
M1$/$2 & 1.39 & (0.244,\ 2.19) \\
NS1$/$2 & 1.78 & (0.585,\ 2.62) \\[5pt]
\multicolumn{2}{@{}l}{\textit{Model---Tridiagonal}} & \\
PB2 & 2.35 & (1.24,\ 3.17) \\
PB1 & 2.52 & (1.29,\ 3.37) \\
PA & 2.08 & (1.03,\ 2.91) \\
$\mathrm{HA}^{\dag}$ & 1.54 & (0.381,\ 2.47) \\
$\mathrm{NP}^{\dag}$ & 2.46 & (1.25,\ 3.28)\\
NA & 2.20 & (1.04,\ 3.04) \\
M1$/$2 & 1.37 & (0.168,\ 2.44) \\
NS1$/$2 & 1.88 & (0.770,\ 2.79) \\
\hline
\end{tabular*}
\end{table}

To more thoroughly unravel the evolutionary history of influenza A, we
need to address the relative level of evolutionary diversity between
segments. To this end we include segment-specific effects whose
posterior estimates we summarize in Table \ref{tab:fixed} for both the
IND and the TRI-P models. We report both the posterior mean and the
95\% BCIs with the range of the TRI-P intervals generally slightly
reduced from the IND, perhaps indicative of the TRI-P being a more
appropriate covariance structure with which to model the correlated
parameters of the different segments. PB1 returns the highest
posterior mean estimates of $\iTMRCA$ which at season 1993 for IND is
$\hat{\iFE}_{\mathrm{IND,PB2}} = 2.51$ (1.30, 3.40) and for
TRI-P is $\hat{\iFE}_{\mathrm{TRI\mbox{-}P,PB2}} = 2.35$ $(1.24, 3.17)$. HA and M1$/$2 yield the lowest posterior means of
$\iTMRCA$ which, also given at season 1993, are $\hat{\iFE}_{\mathrm{IND,HA}} = 1.37$ $(0.161, 2.17)$ and
$\hat{\iFE}_{\mathrm{IND,M1/2}} = 1.39$ $(0.244, 2.19)$
respectively for IND and $\hat{\iFE}_{\mathrm{TRI\mbox{-}P,HA}} =
1.54$ $(0.381, 2.47)$ and $\hat{\iFE}_{\mathrm{TRI\mbox{-}P,M1/2}} = 1.37$ $(0.168, 2.44)$ respectively for TRI-P. This
implies that genetic diversity is maintained longer in PB1 than HA and
M1$/$2 with the difference between these segments for $\iTMRCA$ on the
order of an entire year. We go further than \citet{Rambaut08} who
simply observe the differences in the $\iTMRCA$ of different segments
of influenza A by formally testing whether certain segments maintain
greater diversity than others. One relationship of particular
interest, explored but not formally tested by \citet{Rambaut08},
is the comparison between NP and HA, superscripted in
Table \ref{tab:fixed}. When assuming equal prior probability of both
outcomes (0.5 each), the Bayes factors (BF) of whether the $\iTMRCA$ of
NP is greater than HA are 179 for the IND model and 16.1 for TRI-P.
This means that the posterior probability of NP being greater than HA
in the IND model is 0.994 and 0.941 in the TRI-P, providing strong
support for the hypothesis that NP maintains greater genetic diversity
than HA.


Finally, we approach teasing out the correlation between segments by
estimating the posterior distribution of all possible segment order
permutation within the tridiagonal covariance matrix in the TRI-P
model. The three groupings of \{HA, M1$/$2\}, \{NP, NS1$/$2\}, and \{PA, PB1,
PB2\} originally posited by \citet{Rambaut08} arise from ad hoc
multidimensional scaling results. For notational convenience, we now
refer to these segments by number and place them in alphabetical order
as 1 (HA), 2 (M1$/$2), 3 (NA), 4 (NS1$/$2), 5 (NP), 6 (PA), 7 (PB1), and 8
(PB2). This notational device means that we refer to a model with both
the \{HA, M1$/$2\} and \{PA, PB1, PB2\} grouping as $M_{\{12\}\{678\}}$.
Note that there is no implied ordering of the remaining unlisted
segments so that $M_{\{12\}}$ and $M_{\{678\}}$, say, have some
overlapping groupings. There are seven pairs of neighbors possible in
the tridiagonal matrix because the two segments occupying the corners
of the diagonal of the covariance matrix are restricted to having a
single neighbor and the middle six segments can each have two
neighbors.

We are interested specifically in positive correlation between segments
as we want to clarify which segments have similar evolutionary
histories. With this purview, we focus on results conditional on the
correlation $\rho$ being greater than 0.2; this provision occurs with
approximately 0.593 posterior probability, but eases interpretation.
In general, we assume there are two segments $i$ and $i^{\prime}$
where $i, i^{\prime} \in\{1, \ldots, 8\}$, and $i \neq
i^{\prime}$ which are selected as neighbors. We summarize the
posterior probability of models in which two segments are neighbors for
our subset of interest, $P(M_{\{ii^{\prime}\}} \vert\seqs, \rho>
0.2)$, in Table \ref{tab:permute}. The posterior probability of the
strongly hypothesized \{HA, M1$/$2\} group is very high, $P(M_{\{12\}}
\vert\seqs, \rho> 0.2) = 0.768$. The posterior probability of the
\{PB1, PB2\} pairing is also high, $P(M_{\{78\}} \vert\seqs, \rho>
0.2) = 0.493$. Previously unidentified as a potential pairing is \{NP,
PB2\}, $P(M_{\{48\}} \vert\seqs, \rho> 0.2) = 0.378$ as well as \{NP,
PA\}, $P(M_{\{46\}} \vert\seqs, \rho> 0.2) = 0.313$ which implies that
the NP segment might be just as strongly aligned with the \{PA, PB1,
PB2\} grouping as with NS1$/$2 because the posterior probability of the
\{NP, NS1$/$2\} grouping is similar, $P(M_{\{45\}} \vert\seqs, \rho> 0.2)
= 0.335$.

\begin{table}
\caption{\textup{Posterior probability of segments as neighbors.} The
values represent the posterior probability of two segments having
correlation in the tridiagonal with permutation (TRI-P) model
conditional on the correlation being greater than 0.2 which occurs with
0.593 probability. For notational convenience we refer to the segments
by number, 1 (HA), 2 (M1$/$2), 3 (NA), 4 (NP), 5 (NS), 6 (PA), 7 (PB1),
and 8 (PB2), so the model with HA and M1$/$2 as neighbors is
$M_{\{12\}}$. Hypothesized structural groupings are indicated by the
superscript, \{HA, M1$/$2\} is $\{12\}$, \{NS1$/$2, NP\} is $\{45\}$,
and $\{$PA, PB1, PB2$\}$ is a subset of $\{678\}$. The posterior
probability of \{HA, M1$/$2\} being grouped as neighbors is the
highest $(P(M_{\{12\}} \vert\seqs, \rho> 0.2) = 0.768)$ suggesting
these two segments share similar evolutionary histories}\label{tab:permute}
\begin{tabular*}{\tablewidth}{@{\extracolsep{\fill}}lcd{1.7}d{1.7}d{1.3}d{1.4}d{1.3}ll@{}}
\hline
&\multicolumn{1}{c}{\textbf{PB2}} & \multicolumn{1}{c}{\textbf{PB1}} & \multicolumn{1}{c}{\textbf{PA}} & \multicolumn{1}{c}{\textbf{HA}} & \multicolumn{1}{c}{\textbf{NP}} & \multicolumn{1}{c}{\textbf{NA}} & \multicolumn{1}{c}{\textbf{M1$\bolds{/}$2}} & \multicolumn{1}{c@{}}{\textbf{NS1$\bolds{/}$2}} \\
\hline
\textbf{PB2} & --- & 0.493^{\{78\}} & 0.297^{\{68\}} & 0.181 & 0.378& 0.123 & 0.185 & 0.165 \\
\textbf{PB1} & & \multicolumn{1}{c}{---} & 0.284^{\{67\}} & 0.151 & 0.257 & 0.313 & 0.0990& 0.160 \\
\textbf{PA} & & & \multicolumn{1}{c}{---} & 0.192 & 0.313 & 0.230 & 0.173 & 0.290 \\
\textbf{HA} & & & & \multicolumn{1}{c}{---} & 0.0620 & 0.154 & 0.768$^{\{12\}}$ & 0.157 \\
\textbf{NP} & & & & & \multicolumn{1}{c}{---} & 0.345 & 0.0880 & 0.335$^{\{45\}}$ \\
\textbf{NA} & & & & & & \multicolumn{1}{c}{---} & 0.222 & 0.283 \\
\textbf{M1$\bolds{/}$2} & & & & & & & \multicolumn{1}{c}{---} & 0.304 \\
\textbf{NS1$\bolds{/}$2} & & & & & & & & \multicolumn{1}{c@{}}{---} \\
\hline
\end{tabular*}
\end{table}

Again, we formally test these hypothetical groupings against the null
hypothesis that all permutations of the tridiagonal covariance matrix
are equally likely and summarize the results in Table \ref{tab:BF}. We
make these calculations unconditional on the correlation $\rho$.
Notice in column 2 of Table \ref{tab:BF}, the posterior probability of
grouping \{HA, M1$/$2\} is 0.624 and has a prior probability of 0.0357
leading to a significant BF of 17.5. Again, \{NP, NS1$/$2\} has a weaker
BF of 8.50 lending some doubt to the hypothesis that these two segments
have similar evolutionary histories. Finally, the posterior
probability that all three pairings are found is 0.0741 which, given
all models are thought a priori to have equal probability, has a prior
probability of 0.000600. This leads to a very decisive BF of 124 which
strongly supports Rambaut's et~al. (\citeyear{Rambaut08}) hypothesized groupings of
segments.

\begin{table}[b]
\caption{\textup{Bayes factors (BFs) of hypothesized structural
groupings being nearest neighbors.} BFs are calculated for the
tridiagonal with permutations (TRI-P) model testing the support for the
hypothesized structural groupings of \{HA, M1$/$2\}, \{NS1$/$2, NP\},
and $\{$PA, PB1, PB2$\}$. The posterior odds, prior odds, and BFs are
reported for each grouping individually and then for all three
occurring together. All possibilities of groupings are considered
equally probable for the prior odds. There is strong support for the
\{HA, M1$/$2\} pairing (BF $=$ 17.5) and decisive support for all three
structural groupings being selected as neighbors (BF $=$ 124)}\label{tab:BF}
\begin{tabular*}{\textwidth}{@{\extracolsep{\fill}}ld{2.4}d{1.4}d{2.4}d{3.6}@{}}
\hline
&\multicolumn{1}{c}{\textbf{\{HA,} \textbf{M1$\bolds{/}$2\}}} & \multicolumn{1}{c}{\textbf{\{NP,} \textbf{NS1$\bolds{/}$2\}}} & \multicolumn{1}{c}{\textbf{\{PA, PB1, PB2\}}} & \multicolumn{1}{c@{}}{\textbf{All three}}\\
\hline
Posterior odds & 0.624 & 0.304 & 0.195 & 0.0741 \\
Prior odds & 0.0357 & 0.0357 & 0.0179 & 0.000600 \\
Bayes factor & 17.5 & 8.50 & 10.9 & 124 \\
\hline
\end{tabular*}
\end{table}

\begin{figure}

\includegraphics{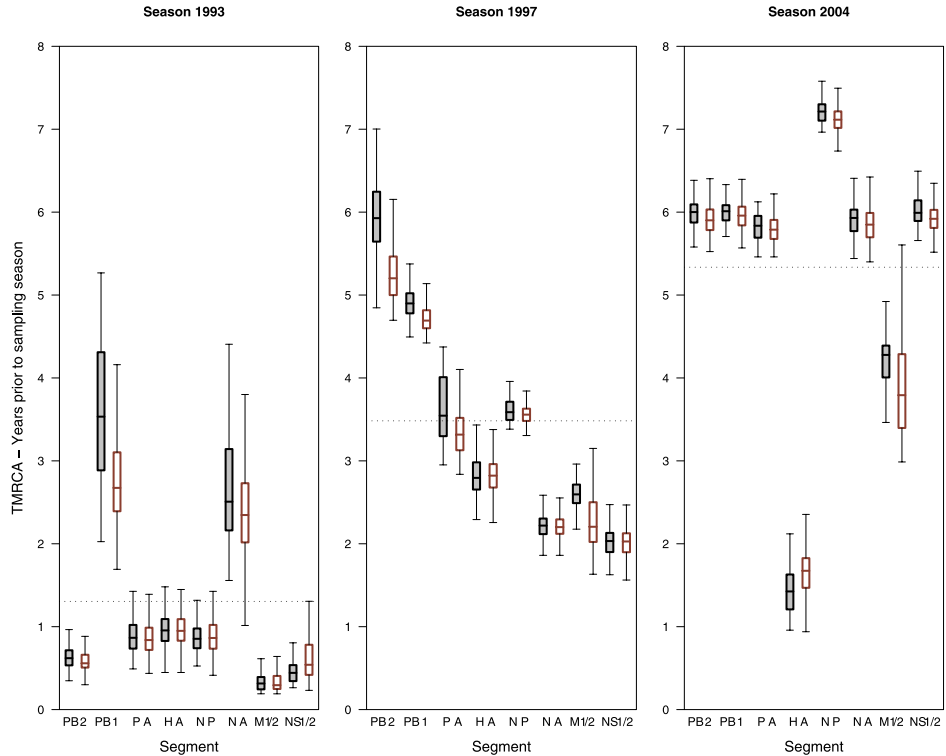}

\caption{\textup{Box plots of stratified and hierarchical estimates for
select seasons.} Box plots of stratified (gray) and hierarchical
(blue) for the selected seasons 1993, 1997, and 2004. On the x-axis
are segments listed from left to right in descending order of size, and
on the y-axis are adjusted TMRCA. Selected seasons clearly display
shrinkage toward the mean of the hierarchical versus stratified
distributions.}\label{fig:box}
\end{figure}

We have already demonstrated the ease with which BF can be assigned to
competing joint hierarchical analysis models once we have recycled the
realizations from stratified analyses into a single statistical
framework. However, an additional advantage is that this hierarchical
posterior distribution demonstrates shrinkage toward the mean when
compared against the stratified results, leading to more sensible
segment-specific estimates. Figure \ref{fig:fitted} displays the
marginal hierarchical posterior distributions superimposed on the
stratified results for all eight segments over time. In
Figure \ref{fig:fitted} we demonstrate that overall the direction of
shrinkage estimates from the stratified mean (black) to the
hierarchical mean (color) draws toward the grand mean for that time.
Shrinkage toward the mean especially tempers outliers, a phenomenon
illustrated in Figure \ref{fig:box}. Focusing on 1993, this season
emits positive outliers in PB1 and NA. Season 1997 has positive
outliers in PB2 and PB1 and 2004 yields a negative outlier in HA and a
positive outlier in NP. In 1993, coverage drastically shifts down
for PB1 and NA as well as shifts up for M1$/$2 and NS1$/$2. In season
1997, coverage greatly decreases and shifts downward for the outlier
PB2 and in season 2004, coverage shifts up for HA and shifts down for
NP. Figure \ref{fig:fitted} also illustrates further advantages of
this technique, including the reduced credible intervals for the
hierarchical distribution relative to the stratified. Finally,
especially apparent in this figure is a cautionary reminder of the wide
variability present in the inappropriately independent, stratified
results. For example in 1993, the stratified estimate uncertainty of
PB1 spans around 3 years whereas that for PB2 spans around 0.5. These
differences in variability mitigate when basing conclusions on the
point estimates of the stratified results. Therefore the advantages of
reusing the stratified analyses in a joint model lie not only in the
ability to assign across-segment BFs and the incorporation of highly
desirable shrinkage estimators that lead to improved estimation
[\citet{Efron77}], but also in enhanced modeling capabilities that more
accurately represent the variability in the segments.

\section{Conclusions}
\label{sec:discuss}
Increasing dataset sizes are engulfing the scientific community
[\citet{Anderson08}] demanding novel approaches to statistical
analysis. While the introduction of GPU programming to the statistical
community promises solutions in the near-future [\citet{Suchard10}],
the daunting task of analyzing these massive datasets is currently made
realistic by partitioning them into smaller, more tractable sizes.
This stratification, while facilitating fast estimation, results in
overparameterization and ignores the correlation between parameters
across strata. Additionally, stratification fails to profit from the
massive amounts of data available because parameters are estimated from
siloed strata, removed from the implicit context that motivated the
initial data collection.

Ideally, given no computational constraints, related and exchangeable
groups are represented by a hierarchical model. This framework
efficiently pools information across groups while accounting for the
correlation between them. This single unified model also makes it easy
to draw dataset-wide inference. Finally, hierarchical models lead to
improved estimators due to shrinkage toward the mean; this well-known
phenomenon is termed Stein's paradox [\citet{Efron77}].

Perhaps a more familiar approach to constructing this full hierarchical
model is sequential Monte Carlo (SMC) [\citet{Doucet01}].
\citet{Chopin02}, \citet{Ridgeway02}, and \citet{Balakrishnan06} use particle
filtering as a SMC solution to the massive data problem, building up
the posterior distribution of the complete data by incrementally
introducing a small number of data into the posterior distribution
using importance sampling [\citet{Cappe07}]. This form of SMC is
highly effective for linearly organized data such as time series
observations but is inappropriate for data divided into exchangeable
groups. Further, particle filters do not completely recycle the
preliminary analyses.

Our methods in this paper create a new strategy, combining the
advantages of stratification, namely speed, with the statistical
framework of hierarchical modeling. Any hypothesis addressed in the
subpar stratified model can be reused, benefiting from assigned
measures of statistical certainty. Our methods capitalize on the
intermediate realizations from stratified analyses, recycling them into
the hierarchical model by reweighting via importance sampling. From
the standpoint of the evolutionary history of influenza A,
\citet{Rambaut08} are ambitious in their goal of understanding
individual segments within the larger context of the complete genome.
Our methods enable us to revisit \citet{Rambaut08}'s conclusions
with the insight afforded by a hierarchical statistical framework.

We find that for these data, the TRI-P model is quite sufficient for
our re-examination of the biological questions. However, in some
circumstances it may be necessary to allow for some small degree of
correlation between segments that are not nearest neighbors. One way
to accommodate this correlation is by assigning a prior inverse-Wishart
distribution to $\cov$. Our approach is to center the inverse-Wishart
distribution on the structured covariance matrix [as in
\citet{Boscardin04}]. The degrees of freedom of the inverse-Wishart
provide a tuning parameter. As the degrees of freedom go to infinity
the extra correlation goes to zero and we recover our original model.
The use of this model requires that we replace the Gibbs sampling step
for $\cov$ with a Metropolis--Hastings step, and so adds some
computational burden.

The applications of reusing, recycling, and reweighting are limited
only by the biological questions of interest. This flexible framework
has far-reaching value into areas such as resequencing and
phylogeography [\citet{Knowles04}], in other words, situations where
computational complexity forces data partitioning, preventing the more
appropriate hierarchical model. From an applied statistician's
perspective, this technique delivers a much needed strategy for
analyzing massive datasets.

\section*{Acknowledgments}
We are thankful to Robert Weiss for modeling discussion and suggestions
and to Eddie Holmes and Andrew Rambaut for generously providing the
results from their analyses and feedback. Interested readers may
request the source-code for this project from J. A. Tom.

\begin{supplement}[id=suppA]
\sname{Supplement A}
\stitle{Details of sampling from the complete model\break}
\slink[url]{http://lib.stat.cmu.edu/aoas/349/supplement.pdf}
\slink[doi]{10.1214/10-AOAS349SUPP} 
\sdatatype{.pdf}
\sdescription{We detail the sampling steps for our complete model
outlined in Section \ref{sec:complete} and our constrained covariance
matrices model outlined in Section \ref{sec:constrained}.}
\end{supplement}

\printaddresses

\end{document}